\documentclass[twocolumn,showpacs,preprintnumbers,amsmath,amssymb]{revtex4}
\usepackage{graphicx}
\usepackage{dcolumn}
\usepackage{bm}

\newcommand{\bef}{\begin{figure}}
\newcommand{\eef}{\end{figure}}

\newcommand{\be}{\begin{equation}}
\newcommand{\ee}{\end{equation}}
\newcommand{\bea}{\begin{eqnarray}}
\newcommand{\eea}{\end{eqnarray}}
\begin{document}

\title{Quark participants and global observables}

\author{Pawan Kumar Netrakanti and Bedangadas Mohanty }
\affiliation{Variable Energy Cyclotron Centre, Kolkata 700064, India}
\date{\today}
\begin{abstract}
We show that the centrality dependence of charged particle and 
photon pseudorapidity density at midrapidity along with the transverse 
energy pseudorapidity density at SPS and RHIC energies scales with the 
number of participating constituent quarks. The number of charged particles 
and transverse energy per participant constituent quark is found to 
increase with increase in beam energy.
\end{abstract}
\pacs{25.75.Ld}
\maketitle

\section{INTRODUCTION}
One of the challenges in relativistic heavy ion collisions is to
measure and study the large number of particles produced in
such reactions. Measurement of particle density and transverse energy
density in rapidity is a convenient way to describe particle production 
in heavy ion collisions. The pseudorapidity density ($dN/d\eta$) and 
transverse energy pseudorapidity density ($dE_{T}/d\eta$) at mid-rapidity
is found to increase with increase in centrality of the reaction
at SPS and RHIC energies~\cite{wa98photon,wa98charge,rhic130,rhic200}. 
This has been claimed to be understood
using a simple geometrical picture of collision. 
At SPS energies, it was found that the particle production scales
with the number of participating nucleons ($N_{N-part}$). 
\begin{equation}
\frac{dN}{d\eta} \propto N_{N-part}^{\alpha}
\end{equation}
The value of $\alpha$ for photons and charged particles at SPS
energies were found to be 1.12 $\pm$ 0.03 and 1.07 $\pm$ 0.05
respectively~\cite{wa98photon,wa98charge}. While that for $E_{T}$
is 1.08 $\pm$ 0.06.
Within the quoted systematic errors the value of
$\alpha$ is similar for both photons, charged particles and transverse
energy. 
The value of $\alpha$ indicates a deviation from the picture of a 
naive wounded nucleus model ($\alpha$ = 1).  

At RHIC energies, it was found that the contribution from hard
processes had a major role in understanding particle 
production~\cite{kharzeev}.
The centrality dependence of charged particle pseudorapidity density 
($dN_{\mathrm ch}/d\eta$)
was explained using the following relation,
\begin{equation}
\frac{dN_{ch}}{d\eta} \propto \beta N_{N-part} + (1 - \beta)N_{N-coll}
\end{equation}
where the parameter $\beta$ is the relative fraction of particles produced 
in soft collisions, and (1 - $\beta$) is the relative fraction produced in 
hard collisions. 
It was observed that the fraction of the hadron multiplicity originating
from hard processes at centre of mass energy ($\sqrt{s}$) 56 GeV was 22\% 
and that 
at 130 GeV was 37\%. However this fraction does not increase much for
200 GeV, thereby bringing in some inconsistency with such an approach. 
This is because one expects the relative contribution from hard process 
to increase with increase in collision energy.

Another approach is to consider that the  nucleus-nucleus collision is 
a superposition of constituent quarks collisions. 
Such a model has been used to
show that the centrality dependence of charged particle pseudorapidity 
density at midrapidity for RHIC energies is proportional to the number
of participating constituent quarks ($N_{q-part}$)~\cite{voloshin}.

In this brief report, we shall address the following aspects using the
constituent quark model. 
(a) Does the above scaling of charged particle pseudorapidity 
density with number of participating constituent quarks holds good for
other global observables like photon pseudorapdity density ($dN_{\mathrm \gamma}/d\eta$) 
and transverse energy pseudorapidity density ($dE_{\mathrm T}/d\eta$) ? 
(b) Is such a scaling also observed at SPS energies ?

In the next section, we briefly describe the calculation of number of
participating nucleons and number of quark participants for various
centre of mass energies. In section~III, we present the results and 
finally we summarize in section~IV.

\bef
\begin{center}
\includegraphics[scale=0.4]{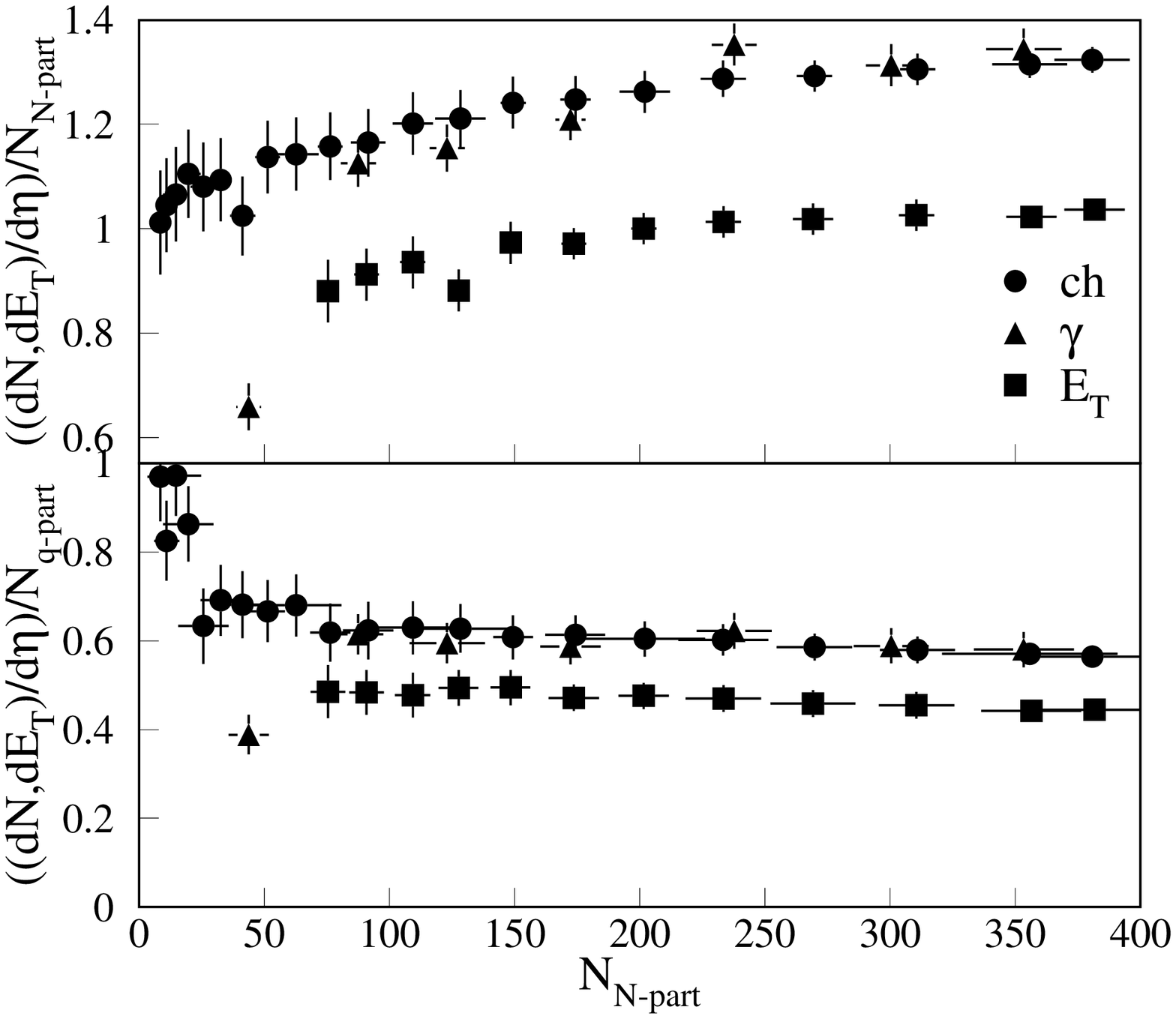}
\caption{$N_{\mathrm ch}$,$N_{\mathrm \gamma}$ and $E_{\mathrm T}$ per nucleon
and quark participant as a function of number of participating nucleons for
SPS energy.
}
\label{fig1}
\end{center}
\eef
\section{Calculation of $N_{N-part}$ and $N_{q-part}$}
The mean number of nucleon and quark participants is calculated 
in a similar manner as in Ref.~\cite{voloshin}. A Wood - Saxon nuclear 
density profile as given below, is used for our calculations.
\begin{equation}
	n_{A}(r) =   \frac{n_{0}}{1+exp[(r-R)/d]},   
\end{equation}
with  parameters $n_{0}$ = 0.17 $fm^{-3}$, 
R=(1.12$A^{1/3} - 0.86A^{-1/3}$)fm, d=0.54fm.
The $N_{N-part}$ for nucleus-nucleus (AB) collisions is calculated using the
relation,
\bea
   N_{part,AB} = \int d^{2}sT_{A}(\vec{s})
\{ 1 - [1-\sigma_{NN}T_{B}(\vec{s}-\vec{b})/B]^{B} \} \\ \nonumber
+ \int d^{2}sT_{B}(\vec{s})
\{1 - [1-\sigma_{NN}T_{A}(\vec{s}-\vec{b})/A]^{A}\}
\eea
where $T(b)~=~\int_{- \infty}^{+ \infty} dz n_{A}( \sqrt{b^{2}+z^{2}})$, 
is the thickness function which is defined as the probability for having a 
nucleon - nucleon (NN) collision within the transverse area element 
db when one nucleon is situated at an impact parameter b relative to another 
nucleon. We use the inelastic NN cross section 
$\sigma_{NN}$ = 30 mb , 41mb , 42mb  
at $\sqrt{s}$ = 17.3 GeV, 130 GeV, 200 GeV  respectively. 
In a similar manner the $N_{q-part}$ is also calculated keeping in mind, the
density was changed to three times that of the nucleon density( $n_{0}^{q}$ 
= 3$n_{0}$ = 0.51 $fm^{-3}$) . The cross sections  are also changed in 
accordance with $\sigma_{qq}$ = $\sigma_{NN}$/9 = 3.33 mb, 4.55 mb, 
4.66 mb for 17.3 GeV, 130 GeV, 200 GeV respectively~\cite{voloshin}. 
The sensitivity of the results to choice of $\sigma_{qq}$ has already 
been discussed in Ref.~\cite{voloshin} hence not done in this work.

\bef
\begin{center}
\includegraphics[scale=0.4]{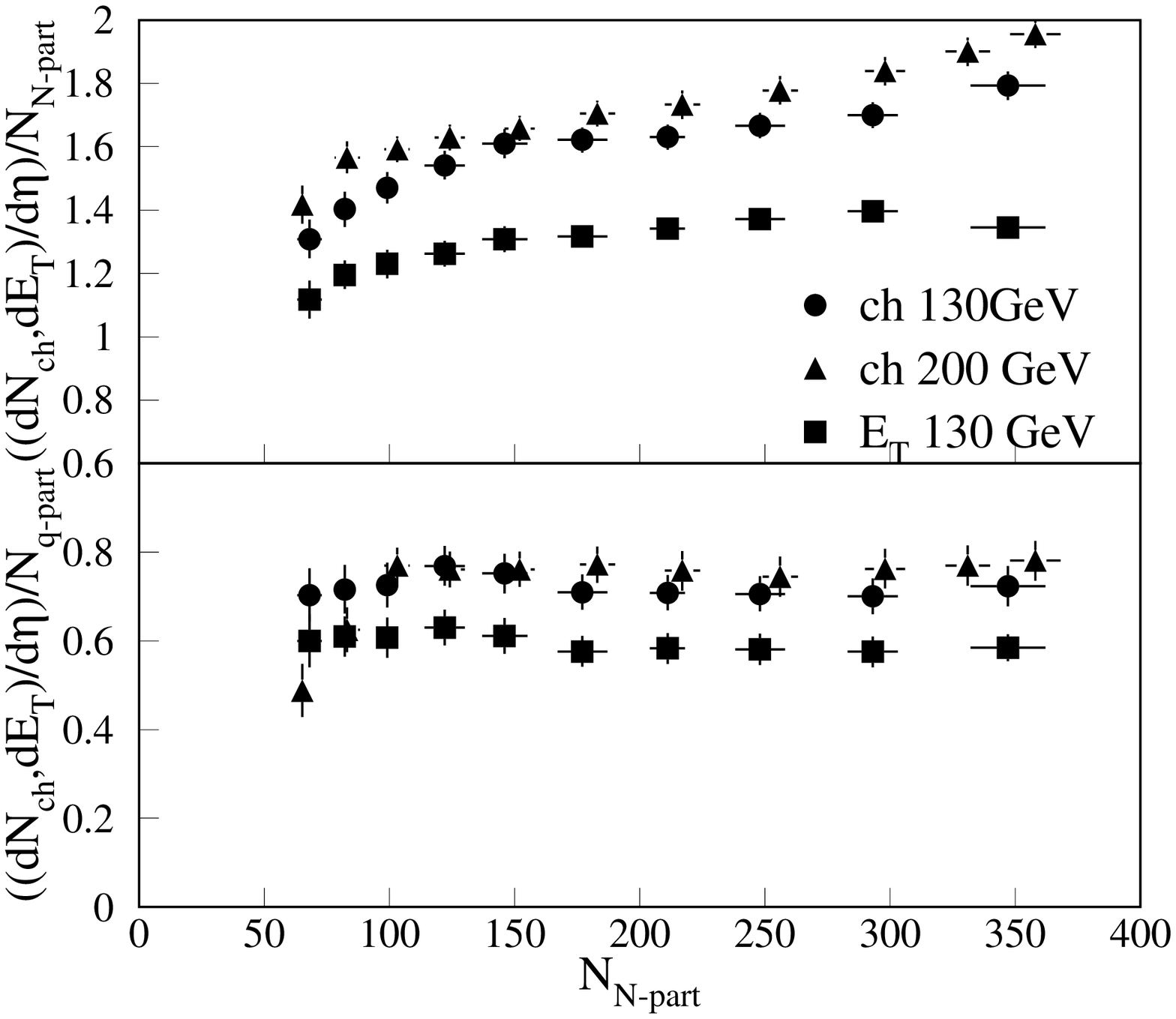}
\caption{$N_{\mathrm ch}$ and $E_{\mathrm T}$ per nucleon
and quark participant as a function of number of participating nucleons
for RHIC energies.
}
\label{fig2}
\end{center}
\eef

\section{Results}
Figure~\ref{fig1} shows the $(dN_{\mathrm ch}/d\eta)/N_{N-part}$,
$(dN_{\mathrm \gamma}/d\eta)/N_{N-part}$ and $(dE_{\mathrm T}/d\eta)/N_{N-part}$
as a function of centrality at SPS energy. 
The lower panel shows the values for
per quark participant. The error bars shown are the systematic errors.
The data values are for the WA98 experiment~\cite{wa98photons,wa98charge}.
It is observed that the values of the observables
per nucleon participant increases as one goes from peripheral collisions
to central collisions. Whereas it remains fairly constant for the case
of quark participants. It may also be noted that the values for charged
particles and photons are of similar order.

Figure~\ref{fig2} shows the $(dN_{\mathrm ch}/d\eta)/N_{N-part}$,
and $(dE_{\mathrm T}/d\eta)/N_{N-part}$ as a function of centrality 
for $\sqrt{s}$ = 130 GeV and $(dN_{\mathrm ch}/d\eta)/N_{N-part}$ for 
$\sqrt{s}$ = 200 GeV at RHIC. The lower panel shows the values for 
per quark participant. The error bars shown are the systematic errors.
The data taken are from the PHENIX~\cite{rhic130} and 
PHOBOS experiments~\cite{rhic200}.
Similar to the case of  SPS energy, here also the 
values of the observables per nucleon participant increases as one goes from 
peripheral collisions to central collisions. While it remains fairly 
constant for the case of quark participants. The differences between
$(dN_{\mathrm ch}/d\eta)/N_{N-part}$ or $(dN_{\mathrm ch}/d\eta)/N_{q-part}$
at $\sqrt{s}$ = 130 GeV and 200 GeV is not much. However there is a general
trend of increase in value of $(dN_{\mathrm ch}/d\eta)/N_{q-part}$ and 
$(dE_{\mathrm T}/d\eta)/N_{q-part}$ with increase in $\sqrt{s}$.

\section{SUMMARY}
We find that the particle multiplicity density and transverse energy
density scales  linearly with the number of constituent quark
participants for SPS and RHIC energies. 

\end{document}